\documentclass[aps,twocolumn,showpacs,showkeys,superscriptaddress,amsmath,amssymb,nofootinbib,floatfix,prc]{revtex4-1}

\usepackage{graphicx}
\usepackage{subfigure}
\usepackage{bm}

\makeatletter

\begin{document}

\title{Forward-backward multiplicity correlations in relativistic heavy-ion collisions in a superposition approach}

\author{Adam Olszewski}
\email{Adam.Olszewski.Fiz@gmail.com} 
\affiliation{Institute of Physics, Jan Kochanowski University, PL-25406~Kielce, Poland}

\author{Wojciech Broniowski}
\email{Wojciech.Broniowski@ifj.edu.pl} 

\affiliation{Institute of Physics, Jan Kochanowski University, PL-25406~Kielce, Poland}
\affiliation{Institut de Physique Th\'eorique, CNRS/URA 2306, F-91191 Gif-sur-Yvette, France}
\affiliation{The H. Niewodnicza\'nski Institute of Nuclear Physics, Polish Academy of Sciences, PL-31342 Krak\'ow, Poland} 

\date{ver. 2, 22 August 2013}

\begin{abstract}
We analyze the multiplicity correlations between distant forward and backward rapidity regions in relativistic heavy-ion collisions in a 
superposition framework, where the particle production occurs through independent emission from correlated sources. 
This in principle allows for inferring information
on the long-range forward-backward correlations of the sources in the earliest phase of the collision, based
solely on the experimental information on the statistical features of the observed particle distributions. 
Our three-stage study incorporates 
the fluctuations of parton production in the early phase, the effect of intermediate hydrodynamic evolution, as well as fluctuations in the production of particles 
at freeze-out. We investigate the dependence of the results on the features of the overlaid distributions and hydrodynamics, as well as the centrality dependence. 
We analyze the existing data from the STAR Collaboration. 
Predictions for the forward-backward multiplicity correlations in Pb+Pb collisions to be analyzed at the LHC are also made. 
\end{abstract}



\maketitle

\section{Introduction \label{sec:intro}}

The forward-backward (F-B) multiplicity correlations in relativistic heavy-ion collisions, recently 
measured at RHIC~\cite{Back:2006id,Tarnowsky:2006nh,Tarnowsky:2008am,Abelev:2009ag}, were followed with a number of theoretical 
studies~\cite{Brogueira:2006yk,Brogueira:2007ub,Haussler:2006rg,Bzdak:2009dr,Bzdak:2009xq,Yan:2010et,Lappi:2009vb,%
Lappi:2010ek,Bialas:2010zb,Bialas:2011bz,Bialas:2011vj,Bialas:2011xk,Bzdak:2011nb,Bzdak:2012tp,Fialkowski:2012ma}. The primary goal 
of these analyses is to get insight into the space-time dynamics of the earlies stages of the reaction, probed via the long-range 
correlations. A simplified 
statistical understanding~\cite{Brogueira:2006yk,Brogueira:2007ub,Bialas:2011bz} of the problem is gained when one considers initially formed {\em sources} 
(wounded nucleons~\cite{Bialas:1976ed,Bialas:2008zza,Broniowski:2007ft}, possibly amended with binary collisions~\cite{Kharzeev:2000ph,Back:2001xy}, Glasma~\cite{Kovner:1995ja,Iancu:2000hn,Lappi:2010cp,Lappi:2011zz}, 
dual strings~\cite{Amelin:1994mf,Brogueira:2006yk}), which later lead to particle production occurring {\em independently} (and with the same distribution) 
from each source. Such {\em superposition models} lead to simple relations 
between the statistical features (moments, correlations) of the distributions of sources and  of the produced particles, which involve properties of the overlaid distribution 
as parameters~\cite{Brogueira:2006yk,Bialas:2011bz}. 

In this paper we analyze in detail the predictions of this framework, tailoring it closely to the popular description of the 
relativistic heavy-ion dynamics based on three stages: initial phase, hydrodynamics, and statistical hadronization~(for a review see, e.g.,~\cite{Florkowski:2010zz}).
The key assumption here is that the emission from a source is universal, i.e., occurs with the same statistical distribution 
independently of the centrality class of the collision.  
This leads to very simple formulas for the statistical measures, in particular for the correlation coefficient 
between the numbers of particles produced in the forward and backward rapidity bins.
Parameters in these relations depend on the features of the overlaid statistical distributions and the properties of hydrodynamics. 

We use the derived relations in two ways. First, using the experimental data from the STAR Collaboration~\cite{Tarnowsky:2006nh,Tarnowsky:2008am,Abelev:2009ag} 
for the correlation and the scaled variance, we obtain an expression for the correlation of sources in the early phase, involving one free parameter 
dependent on the unknown features of the overlaid distributions and hydrodynamics. This straightforwardly generalizes the treatment 
of Brogueira and Dias de Deus~\cite{Brogueira:2006yk} and Bia\l{}as, Bzdak, and Zalewski~\cite{Bialas:2011bz}, who use a Poisson distribution in the two-step approach.
We carry out our analysis for the Au+Au and Cu+Cu collisions at $\sqrt{s_{NN}}=200~{\rm GeV}$, where the data are available, using the 
methodology of Refs.~\cite{Lappi:2009vb,Bzdak:2011nb}.

Second, we use the relations in the opposite direction, starting from the properties of the source distribution and computing the 
F-B correlation of the produced particles. We use the Glauber framework for this purpose. Predictions, involving one free parameter, 
for the Pb+Pb collisions at $\sqrt{s_{NN}}=2.76~{\rm TeV}$ at the Large Hadron Collider (LHC) are made. We predict
a decrease of the F-B correlations of particles with centrality, which follows from the 
decrease of the scaled variance of the sources. The future data from the LHC will verify this scenario. 

The outline of the paper is as follows: In section (\ref{sec:3stage}) we briefly recall the three-stage approach, 
consisting of the early production, the hydrodynamic evolution, and the final freezeout. We then derive the basic relations
linking the statistical features of the sources and of the final particle distributions. 
In Sec.~(\ref{sec:STAR}) we use the data from the STAR Collaboration for the Au+Au and Cu+Cu collisions to obtain the F-B multiplicity correlations of the sources. 
We take an effort to incorporate the intricacies of the STAR measurement, as explained in Ref.~\cite{Lappi:2009vb,Bzdak:2011nb}.
Finally, in Sec.~(\ref{sec:LHC}) we present our predictions for the F-B correlations of particle multiplicities to be experimentally analyzed with the LHC for the Pb+Pb system. 
We use the approach introduced in Sec.~(\ref{sec:3stage}) and the distributions of sources obtained with the mixed model~\cite{Kharzeev:2000ph,Back:2001xy} 
from GLISSANDO~\cite{Broniowski:2007nz}.

\section{Three-stage approach \label{sec:3stage}}

Much of success in the description of the relativistic heavy-ion dynamics has been achieved in a three-stage model 
(for a review see, e.g.,~\cite{Florkowski:2010zz}), consisting of 

\begin{enumerate}
 \item Early production, modeled in terms of the Glauber approach~\cite{Czyz:1969jg,Bialas:1976ed,Bialas:2008zza,Kharzeev:2000ph,Back:2001xy,Broniowski:2007ft}, Glasma~\cite{Kovner:1995ja,Iancu:2000hn,Lappi:2010cp,Lappi:2011zz}, string formation~\cite{Amelin:1994mf,Brogueira:2006yk}, etc.
 \item Hydrodynamic evolution, starting from the initial condition provided by stage~1 (for reviews see, e.g., \cite{Kolb:2003dz,Huovinen:2006jp,Florkowski:2010zz}, and for the 
recent event-by-event studies~\cite{Andrade:2006yh,Werner:2009fa,Petersen:2010cw,Holopainen:2010gz,Gardim:2011xv,Bozek:2011if,%
Schenke:2010rr,Qiu:2011fi,Chaudhuri:2011pa,Bozek:2012en}).
 \item Statistical hadronization, carried out at freezeout right after the hydrodynamic phase ends, 
e.g.,~\cite{Cooper:1974mv,BraunMunzinger:2001ip,Broniowski:2001we,Torrieri:2002jp,Rafelski:2003ju,Becattini:2003wp,Torrieri:2004zz,Wheaton:2004qb,%
Kisiel:2005hn,Amelin:2006qe,Tomasik:2008fq,Chojnacki:2011hb}.
\end{enumerate}

The approach leads to a proper description of such measured quantities as 
multiplicities, spectra, harmonic flow coefficients, femtoscopic properties, etc.,~\cite{Broniowski:2008vp}, thus is viewed as 
a practical framework.
Statistical fluctuations are generated in phases 1 and 3, while phase 2 is assumed to be deterministic (although one may include 
fluctuations also in this phase~\cite{Kapusta:2011gt}). Of course, the most interesting 
are the fluctuations in the initial stage, as they refer to the important physics at the earliest times and may help to 
discriminate between various physics approaches, while the fluctuations at hadronization form a ``statistical noise'' which should be 
disposed of.
While each of the stages is physically involved, including numerous physical parameters, and needless to say, takes a 
huge effort to simulate numerically, certain statistical aspects, as we shall see, can be understood and classified in rather
simple terms, displaying the possible scenarios in the F-B correlations of the earliest phase.

Throughout this paper we use the generic 
concept of {\em sources}, which may be viewed statistically as the density of partons (or fields) in the initial 
phase, turning into the entropy density of the fluid cell in the hydrodynamic phase, which in the end, at freeze-out, gives birth to the observed hadrons
streaming to the detectors.   
In the context of the F-B multiplicity fluctuations, an important assumption is the sufficient kinematic separation of the forward (F) and backward (B) 
rapidity windows, such that the particles produced in from the F source do not fall into the B window, and vice versa. This allows us to trace 
the evolution of a cell with fluctuating sources. Suppose in the early-production phase we separate the F and B cells in spatial rapidity, 
with the original number of sources $s$ denoted as $s_F$ and $s_B$, respectively. 
These numbers fluctuate event-by-event and may be correlated, which is what we 
eventually want to assess. In the initial production mechanism these sources produce partons $p$, whose density supplies 
the initial condition for hydrodynamics.

Assuming that the production occurs from each source in the same manner, we have 
\begin{eqnarray}
p_A= \sum_{i=1}^{s_A} \mu_i, \;\; A=F,B, \label{eq:pA} 
\end{eqnarray}
where the random variable $\mu_i$ is the number of partons produced from the source $i$. 
As mentioned, we assume that the distribution of $\mu_i$ is universal, i.e., does not depend on the location of the cell, and that the production 
from different cells is
{\em independent} from one another. Then the formulas for the superposition model follow (see Appendix~\ref{sec:super}):
\begin{eqnarray}
\langle p_A \rangle &=&  \langle \mu \rangle \langle s_A \rangle, \nonumber \\
{\rm var}(p_A) &=& {\rm var}(\mu) \langle s_A \rangle + \langle \mu \rangle^2 {\rm var}(s_A), \label{eq:pstat} \\ 
{\rm cov}(p_F,p_B) &=& \langle \mu \rangle^2 {\rm cov}(s_F,s_B),  \nonumber
\end{eqnarray}
where $A=F,B$.

The sources $p$ yield the entropy density which constitutes an event-by-event initial condition 
for the collective evolution via hydrodynamics. The hydrodynamic evolution is dynamically complicated, however,
it is deterministic. Thus the evolution of the cell with initially $p$ sources yields $h$ sources at freeze-out 
($p$ and $h$ may be thought of as entropy contained in the cell),
where $h$ is a function of $p$ (Strictly speaking, the hydrodynamic evolution depends not only on the number of sources $p$, but also on their 
spatial distribution, hence the same values of $p$ may lead to somewhat different $h$. Such nuances can only be 
included in a fully numerical simulation. We do not expect them to be relevant for our analysis). 
The phase-space location of the fluid cells is evolved due to the hydrodynamic push, however, the separation 
between the far-lying F and B regions is maintained.  
If the event-by-event fluctuations are not too large, we may expand 
\begin{eqnarray}
 h = t_0  \langle p \rangle + t_1 (p - \langle p \rangle ) + {\cal O}\left ( (p - \langle p \rangle )^2 \right ), \label{eq:hs}
\end{eqnarray}
where $t_i$ are parameters depending on dynamical properties of hydrodynamics.
The higher-order terms may be dropped if $p$ is sufficiently close to $\langle p \rangle$. 
The constant term is written in the form $t_0  \langle p \rangle $. Then
\begin{eqnarray}
\langle h_A \rangle &=& t_0  \langle p_A \rangle, \nonumber \\
{\rm var}(h_A) &=& t_1^2 {\rm var}(p_A), \label{eq:hstat} \\
{\rm cov}(h_F,h_B) &=& t_1^2 {\rm cov}(p_F,p_B). \nonumber
\end{eqnarray}

Equation (\ref{eq:hs}) requires some discussion. If the response of hydrodynamics is such that $h \sim t$, i.e. the entropy in the considered cell 
after the evolution is proportional to the initial entropy, then $t_0 \simeq t_1$. The success of hydrodynamics in reproducing the rapidity spectra 
starting from an initial Glauber condition~(see, e.g.,~\cite{Bozek:2011ua}) supports this scenario. Moreover, it suggests that $t_0$ and $t_1$ are independent 
of centrality in the range of applicability of hydrodynamics. 
Secondly, since in viscous hydrodynamics the entropy is produced, 
we have in a given cell $\langle h \rangle  > \langle p \rangle$, which implies $t_0>1$. The production of entropy depends on the properties of hydrodynamics, but is 
not very large~\cite{Bozek:2011ua} (25-50\%), such that we expect $t_0$ to be only somewhat larger than 1. To summarize, we arrive at the estimates
\begin{eqnarray}
&& t_0 \simeq t_1 \sim 1, \nonumber \\
&& t_0>1. \label{eq:hydexp}
\end{eqnarray}
The parameters $t_0$ and $t_1$ should be treated as specific to a given reaction or rapidity bin, but independent of centrality. 

Finally, statistical hadronization is carried out at freezeout. We assume that a given cell emits $n$ hadrons into a region of phase-space with some 
statistical distribution superimposed over $h$. Similarly to the analogous mechanism in the initial phase, each of the $h$ sources emits independently 
$m$ hadrons with the same distribution. In reality some mixing may occur and particles may be emitted from different cells into the same kinematic region.
However, if the F and B cells are well separated, this effect is negligible. Thus we have  
\begin{eqnarray}
n_A= \sum_{i=1}^{h_A} m_i,  \label{eq:nA} 
\end{eqnarray}
and
\begin{eqnarray}
\langle n_A \rangle &=&  \langle m \rangle \langle h_A \rangle, \nonumber \\
{\rm var}(n_A) &=& {\rm var}(m) \langle h_A \rangle + \langle m \rangle^2 {\rm var}(h_A). \label{eq:hnstat} \\
{\rm cov}(n_F,n_B) &=&  \langle m \rangle^2 {\rm cov}(h_F,h_B). \nonumber
\end{eqnarray}
The three-stage model and the notation introduced above may be summarized with the following diagram:
\[
s \stackrel{\rm init. production}{\longrightarrow} p \stackrel{\rm hydro}{\longrightarrow} h \stackrel{\rm hadronization}{\longrightarrow} n 
\]

Joining Eqs.~(\ref{eq:pstat}-\ref{eq:hnstat}) yields
\begin{eqnarray}
\langle n_A \rangle &=& \alpha \langle s_A \rangle, \nonumber \\
{\rm var}(n_A) &=&  \beta \langle s_A \rangle + \gamma {\rm var}(s_A) ,\label{eq:hnstat2} \\
{\rm cov}(n_F,n_B) &=& \gamma {\rm cov}(s_F,s_B). \nonumber 
\end{eqnarray}
where the combinations of constants are
\begin{eqnarray}
\alpha &=& t_0 \langle \mu \rangle \langle m \rangle, \nonumber \\
\beta  &=& t_0 \langle \mu \rangle {\rm var} (m) + t_1^2 \langle m \rangle^2 {\rm var} (\mu), \label{eq:abg} \\
\gamma &=& t_1^2 \langle \mu \rangle^2 \langle m \rangle^2. \nonumber
\end{eqnarray}
It is convenient to introduce the scaled variance $\omega(x_A) = {\rm var}(x_A)/\langle x_A \rangle$ and the correlation
coefficients $\rho(x_F,x_B)={\rm cov}(x_F,x_B)/(\sigma(x_F)\sigma(x_B))$. Then, for symmetrically separated bins where 
$\sigma(x_F)=\sigma(x_B)$, we can write the relations
\begin{eqnarray}
\omega(n_A)   &=& \delta + \kappa \omega(s_A),  \nonumber \\
\rho(n_F,n_B) &=& \frac{\rho(s_F,s_B)}{1+ \lambda / \omega(s_A)}, \label{eq:omr}
\end{eqnarray}
with
\begin{eqnarray}
\delta &=& \beta/\alpha = \omega(m) +\frac{t_1^2}{t_0} \langle m \rangle \omega(\mu), \nonumber \\
\kappa &=& \gamma/ \alpha = \frac{t_1^2}{t_0} \langle \mu \rangle \langle m \rangle, \label{eq:kap} \\
\lambda &=& \beta/\gamma= \frac{t_0 \omega(m)}{t_1^2 \langle \mu \rangle\langle m \rangle} + \frac{\omega(\mu)}{\langle \mu \rangle}. \nonumber 
\end{eqnarray}

Formulas (\ref{eq:hnstat2},\ref{eq:omr}) express the statistical properties of the event-by-event distributions of the produced 
particles in the F and B bins via the properties of the distribution of the original sources. 
Note that the relation between the correlation coefficients depends on a single combination of 
the unknown parameters of the overlaid distributions and hydrodynamics represented in 
$\lambda$, which makes the qualitative analysis simple. This feature is a derivative of the form of Eq.~(\ref{eq:hs}).

We may also write reversed formulas expressing the properties of the initial sources through the properties of the final particle distributions:
\begin{eqnarray}
\langle{s_A}\rangle &=& \frac{1}{\alpha} \langle n_A \rangle, \nonumber \\
\omega(s_A) &=& \frac{\omega(n_A)}{\kappa} - \lambda, \label{eq:omeg}  \\
\rho(s_F,s_B) &=& \frac{\rho(n_F,n_B)}{1-\delta/\omega(n_A)}. \nonumber  
\end{eqnarray}
For the case of asymmetric bins we have
\begin{eqnarray}
\rho(s_X,s_Y) &=& \frac{\rho(n_X,n_Y)}{\sqrt{1-\delta/\omega(n_X)}\sqrt{1-\delta/\omega(n_Y)}}. \label{eq:ras}
\end{eqnarray}

A feature following from the assumptions in our analysis is the independence of the parameters $\alpha$, $\beta$, $\gamma$, $\delta$, $\kappa$, and $\lambda$ 
on the centrality class. This dependence resides entirely in the statistical properties of the distributions of $s$ or $n$. 

Finally, we note that the inclusion of finite experimental acceptance $a$ amounts to overlaying yet another 
statistical distribution over $n$, with mean $a$ and variance $a(1-a)$. As a result, the parameters $\alpha$, $\beta$, 
and $\gamma$ are changed, however, the form of Eqs.~(\ref{eq:hnstat2}-\ref{eq:omeg})
remains unaltered.

The basic methodology is as follows: Eq.~(\ref{eq:omeg}) involves three independent parameters: $\alpha$, $\kappa$, and $\lambda$ (note that $\delta=\kappa \lambda$).
Thus, knowing from the experiment  $\langle n_A \rangle$, $\omega(n_A)$, and $\rho(n_F,n_B)$ and from the model  $\langle s_A \rangle$, $\omega(s_A)$, and $\rho(s_F,s_B)$
(at a given centrality) allows us to solve the equations for $\alpha$, $\kappa$, and $\lambda$. Comparing the values at various centralities serves as a consistency 
check (more appropriately, one should simultaneously solve the equations at all centralities in the $\chi^2$ sense). 
For the method to be feasible, however, one needs the complete experimental data involving $\langle n_A \rangle$, $\omega(n_A)$, and $\rho(n_F,n_B)$ at all 
centralities, which, unfortunately, is not the case of Refs.~\cite{Tarnowsky:2008am,Abelev:2009ag}, where $\omega(n_A)$ is provided only for two 
centrality classes. For that reason the above program is carried out only partially in the following Sections.
On the model side, we need $\langle s_A \rangle$, $\omega(s_A)$, and $\rho(s_F,s_B)$.

\section{Analysis of the STAR data \label{sec:STAR}} 

As originally pointed out by Lappi and McLerran~\cite{Lappi:2009vb}, statistical interpretation of the STAR 
measurement~\cite{Tarnowsky:2008am,Abelev:2009ag} is affected by correlations to the {\em reference bin} used 
to determine centrality (for large F-B separations, the reference bin takes $|\eta|<0.5$). 
The STAR analysis first  sets the multiplicity in the reference bin, $n_R$, end then with  
this multiplicity fixed computes the variances and correlations for the $F$ and $B$ bins. Finally, averaging over 
$n_R$ is performed in a given centrality class. This procedure
leads to important corrections~\cite{Lappi:2009vb,Bzdak:2011nb} over the naive interpretation of the data. Let us denote the 
correlation between the peripheral (F,B) and central (R) bins as (we consider the symmetric case in $\eta$ all over) 
$\rho(n_F,n_R)=\rho(n_B,n_R)=\rho(n_A,n_R)\equiv R$, where $A=F,B$.
Then the correlations reported by the STAR Collaboration, denoted here as $\rho^*(n_F,n_B)$, relates to the ``true'' 
F-B correlation $\rho(n_F,n_B)$ in 
the following way~\cite{Lappi:2009vb}:
\begin{eqnarray}
\rho^*(n_F,n_B)=\frac{\rho(n_F,n_B)-R^2}{1-R^2}. \label{eq:Rf} 
\end{eqnarray}
As shown by Bzdak~\cite{Bzdak:2011nb}, formula (\ref{eq:Rf}) may be straightforwardly obtained by incorporating the linear 
dependence
\begin{eqnarray}
\langle n_A \rangle_{n_R}=c_0+c_1 n_R, 
\end{eqnarray}
(see Eq.~(5-9) of Ref.~\cite{Bzdak:2011nb} for the derivation). 
Then 
\begin{eqnarray}
c_0=\langle n_A \rangle - \langle n_R \rangle R \frac{\sigma(n_A)}{\sigma(n_R)}, \;\; c_1= R \frac{\sigma(n_A)}{\sigma(n_R)}, 
\end{eqnarray}
where $\sigma(.)$ denotes the standard deviation.
Since experimentally $c_0 \simeq 0$~\cite{Tarnowsky:2008am}, we obtain
\begin{eqnarray}
\omega(n_R)= \frac{\langle n_R \rangle}{\langle n_A \rangle} R^2 \omega(n_A). \label{eq:nre2}
\end{eqnarray}
It also follows~\cite{Bzdak:2011nb} that the scaled variance measured according to the STAR method relates to the usual scaled variance in the following way: 
\begin{eqnarray}
\omega^*(n_A)=\omega(n_A)(1-R^2). \label{eq:Rom}
\end{eqnarray}

\begin{table}[tb]
\caption{Values of the measured parameters used in our analysis. \label{tab:obs}}
\begin{tabular}{|l|cc|}
\hline
                   &  $\rho^\ast(n_F,n_B)$ & $\omega^\ast(n_A)$ \\ \hline
Au+Au, $c=0-10\%$  &  0.58                 & 3.9 \\
Au+Au, $c=40-50\%$ &  0                    & 1.05 \\ \hline
Cu+Cu, $c=0-10\%$  &  0.48                 & 2.7 \\ 
Cu+Cu, $c=40-50\%$ &  0                    & 1.1 \\
\hline
\end{tabular}
\end{table}

With the help of Eq.~(\ref{eq:Rf}) and (\ref{eq:Rom}) we can now write down the relation
\begin{eqnarray}
&& R^2=\rho(s_F,s_B)\left [1-\frac{\delta (1-R^2)}{\omega^*(n_A)}\right] - (1-R^2)\rho^*(n_F,n_B). \nonumber \\
\label{eq:set1} 
\end{eqnarray}
Next, we use Eqs.~(\ref{eq:ras}) and (\ref{eq:nre2}) to obtain the formula
\begin{eqnarray}
R^2=\rho(s_A,s_R)^2\left[ 1\!-\!\frac{\delta (1\!-\!R^2)}{\omega^\ast(n_A)} \right] 
\left[ 1\!-\!\frac{\delta (1\!-\!R^2)}{\omega^\ast(n_A)} \frac{\langle n_A \rangle}{R^2 \langle n_R \rangle}\right]. \nonumber \\ \label{eq:set2}
\end{eqnarray}
We can solve Eq.~(\ref{eq:set1}) for $R^2$, 
\begin{eqnarray}
R^2= \frac{\rho^\ast(n_F,n_B)-[1-\delta/\omega^\ast(n_A)]\rho(s_F,s_B)}{\rho^\ast(n_F,n_B)-1+\delta/\omega^\ast(n_A) \rho(s_F,s_B)},
\end{eqnarray}
plug it to Eq.~(\ref{eq:set2}), and obtain a relation involving the correlations of sources between 
the central (reference) and peripheral bins, $\rho(s_A,s_R)$, the forward and backward bins, $\rho(s_F,s_B)$:
\begin{widetext}
\begin{eqnarray}
\rho(s_A,s_R)^2=\frac{ \left\{ \left[1-\frac{\delta}{\omega^\ast(n_A)}\right]\rho(s_F,s_B)-\rho^\ast(n_F,n_B) \right \}^2}
{\left\{1-\rho^\ast(n_F,n_B)-\frac{\delta}{\omega^\ast(n_A)}\right\}
\left\{\rho(s_F,s_B)-\rho^\ast(n_F,n_B)-\frac{\delta}{\omega^\ast(n_A)} 
\left[ \frac{\langle n_A \rangle }{\langle n_R \rangle} (\rho(s_F,s_B)-1)+\rho(s_F,s_B)\right]\right\} }.
\nonumber \\ \label{eq:long}
\end{eqnarray}
\end{widetext}
We note that the condition $R^2 \le 1$ leads to a limit on the $\delta$ parameter,
\begin{eqnarray}
\delta/\omega^\ast(n_A) \le 1-\rho^\ast(n_F,n_B). \label{eq:cli}
\end{eqnarray}

In formula (\ref{eq:long}) $\delta$ is a model parameter (independent of centrality), cf.~Eq.(\ref{eq:kap}), $\rho^\ast(n_F,n_B)$ and $\omega^\ast(n_A)$ are obtained from the 
published data~\cite{Tarnowsky:2006nh,Tarnowsky:2008am,Abelev:2009ag}, while the ratio $\langle n_A \rangle / \langle n_R \rangle$ reflects 
the acceptance in the forward and central bins. As the central reference bin, extending from $\eta=-0.5$ to $\eta=0.5$, is 5 times 
wider than the peripheral bins, and the pseudorapidity spectra are rather flat in the experimentally covered region, we take
$\langle n_A \rangle / \langle n_R \rangle=0.2$. The two centrality classes for which $\omega^\ast(n_A)$ is available are $c=0-10\%$ and $c=40-50\%$. 
The values of the observables used in our 
analysis are collected in Table~\ref{tab:obs}. These values correspond to the rapidity separations of the forward and backward bins from $\Delta \eta=1.2$ to $\Delta \eta=1.8$, where 
the experimental values do not differ much.

\begin{figure}
\begin{center}
\includegraphics[width=0.45\textwidth]{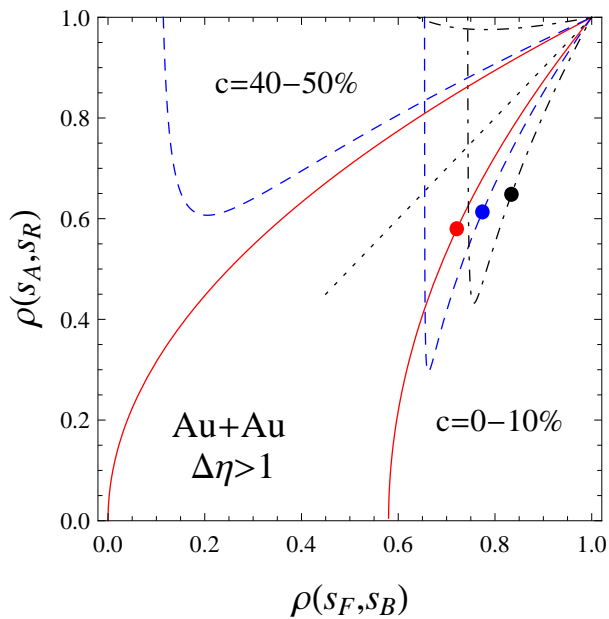}
\caption{(Color online)
Relation (\ref{eq:long}) between the forward-backward correlation of sources $\rho(s_F,s_B)$ and the forward-central correlation of sources $\rho(s_A,s_R)$ 
for the Au+Au collisions and $\Delta \eta > 1$. The solid line is for $\delta=0$ (no superposition), the dashed line for $\delta=0.4$, and the dot-dashed line 
for $\delta=0.8$. The lower three cores correspond to the central collisions, $c=0-10\%$, while the upper three curves to the peripheral collisions, $c=40-50\%$.
The dots indicate the estimate for $\rho(n_f,n_B)\simeq 0.72$ obtained if Ref.~\cite{Bzdak:2011nb}. The dotted line separates the regions $\rho(s_F,s_B)>\rho(s_A,s_R)$ and 
$\rho(s_F,s_B)<\rho(s_A,s_R)$.
\label{fig:fAu}}
\end{center}
\end{figure}

\begin{figure}
\begin{center}
\includegraphics[width=0.45\textwidth]{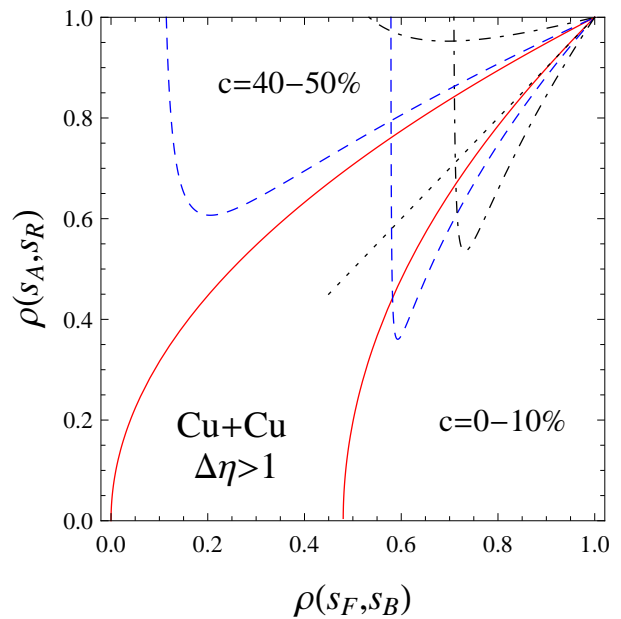}
\caption{(Color online) Same as Fig.~\ref{fig:fAu} for the Cu+Cu collisions.
\label{fig:fCu}}
\end{center}
\end{figure}

The results following from Eq.~(\ref{eq:long}) with the values taken from Table~\ref{tab:obs} are plotted in Figs.~\ref{fig:fAu} and \ref{fig:fCu} for several 
values of the model parameter $\delta$. Each figure contains six curves: three lower ones for $c=0-10\%$ and three higher ones for $c=40-50\%$. The
regions $\rho(s_F,s_B)>\rho(s_A,s_R)$ and $\rho(s_F,s_B)<\rho(s_A,s_R)$ are separated with dotted lines. 
For the formal case $\delta=0$ (no superimposed distributions) we recover $\rho(s_F,s_B)=\rho(n_F,n_B)$ and $\rho(s_A,s_R)=\rho(n_A,n_R)=R$ (solid lines). 
The dashed lines are for $\delta=0.4$, and the dot-dashed for $\delta=0.8$ (significantly higher values are precluded by condition~(\ref{eq:cli})). 
We note that the dashed and dot-dashed curves behave non-monotonically. The sections of the curves which are decreasing are not physical, as they would mean that 
increased forward-backward correlation leads to decreased forward central correlation. The rising parts of the curves are acceptable. We note that for the 
central collisions,  $c=0-10\%$, the rising parts of the curves are in the region $\rho(s_F,s_B)>\rho(s_A,s_R)$. As already stressed in 
Refs.~\cite{Lappi:2009vb,Bzdak:2011nb} (for the case $\delta=0$), this is a puzzling result, meaning that more separated bins are more correlated. 
Our calculation in the superposition model supports this conclusion.
On the other hand, we find that for the peripheral case, $c=40-50\%$, the natural hierarchy is restored, with $\rho(s_F,s_B)<\rho(s_A,s_R)$.

In Ref.~\cite{Bzdak:2011nb}, based on the STAR data published 
in Refs.~\cite{Abelev:2009ag,Tarnowsky:2008am,Abelev:2008ab}, an estimate $\rho(n_F,n_B) \simeq 0.72$ for the Au+Au collisions is made for the case 
of the most central events. We indicate this special value of $\rho(n_F,n_B)$ by the blobs on the curves in Figs.~\ref{fig:fAu} and \ref{fig:fCu}.
With this constraint taken into account, we find that $\rho(s_F,s_B)$ is limited from below by 0.72 and from above by about 0.85 (this limit follows from 
the constraint (\ref{eq:cli}), precluding $\delta$ to be too large). The corresponding values of $\rho(s_A,s_R)$ are between 0.58 and 
0.65.

We recall that the first measurements of the F-B multiplicity correlations at RHIC were carried out by the PHOBOS 
collaboration~\cite{Back:2006id}. Since the statistical measure used in that work is different form the 
correlation coefficient used in this work, we do not use the PHOBOS data in the present analysis.

\section{Predictions for the LHC \label{sec:LHC}}

\begin{figure}
\begin{center}
\includegraphics[width=0.45\textwidth]{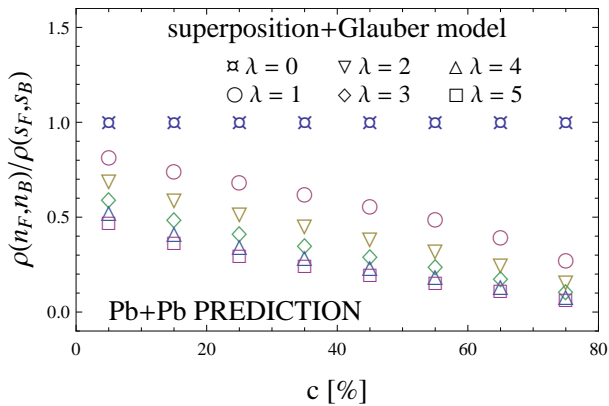}
\caption{(Color online)
Predictions for the ratio of the F-B multiplicity correlation of measured particles to the F-B correlation of sources, plotted as a function of centrality, 
obtained from Eq.~(\ref{eq:omr}) at different values of $\lambda$. The scaled variance $\omega(s_A)$ is taken from GLISSANDO for the mixed model. 
The simulation is for the Pb+Pb collisions at the LHC energy of $\sqrt{s_{NN}}=2.76~{\rm TeV}$. }
\label{f_kr2}
\end{center}
\end{figure}

Finally, we use Eq.~(\ref{eq:omr}) to predict the F-B correlation for the Pb+Pb collisions at the collision energy of
$\sqrt{s_{NN}}=2.76~{\rm TeV}$ at the LHC.\footnote{We assume here that these correlations will 
be measured without correlating to the reference bin, as was the case of the STAR analysis.} The values 
of $\omega(s_A)$ are obtained from the mixed Glauber model as implemented in GLISSANDO. In particular, we take 
the mixing parameter $\alpha=15\%$ and $\sigma_{NN}=67$~mb. 
Figure~(\ref{f_kr2}) shows our predictions, which will allow to extract $\rho(s_F,s_B)$ when the LHC data for the F-B multiplicity 
correlations are analyzed. We note a graduate fall-off of $\rho(n_F,n_B)$ with centrality.
This fall-off is due to the fact that the scaled variance $\omega(s_A)$ decreases with growing centrality. 
The expected value of the the $\lambda$ parameter is in the range $1-2$.

\section{Conclusions}

In this paper we have extended the analysis of correlations in superposition models, previously made in 
Refs.~\cite{Brogueira:2006yk,Brogueira:2007ub,Bialas:2011bz}, to the case of the three-stage approach consisting of 
the early production, hydrodynamics, and statistical hadronization. Simple formulas, linking the statistical 
properties of the F-B correlations in the data and in the original sources have been derived. 
The effect of hydrodynamics may be, under reasonable assumptions, incorporated in terms of just two parameters independent of 
centrality. The relations between the F-B multiplicity correlations of the produced particles and the initial sources 
involve a single parameter, which collects the features of the overlaid distributions and hydrodynamics. 
These one-parameter formulas allow in principle to 
verify the model of the early-phase production with the experiment, under the proviso that the data are sufficiently complete in providing,
for each centrality, not only the F-B correlation coefficient, but also the variance of the number of particles in the F and B bins and their 
multiplicities. 

We have applied the approach on the existent STAR data~\cite{Tarnowsky:2008am,Abelev:2009ag} for the F-B multiplicity correlations,
taking into account the complication explained in Refs.~\cite{Lappi:2009vb,Bzdak:2011nb} which introduces the correlations to the reference bin 
into the framework. Our study confirms the results of Refs.~\cite{Lappi:2009vb,Bzdak:2011nb}, namely, that for the central collisions 
the F-B source multiplicity correlations are stronger than the correlations of the peripheral to central bin,
$\rho(n_F,n_B)>\rho(n_A,n_R)$.

We argue that the complete data, consisting  of 
the average multiplicity and variance in the forward and backward rapidity bins, as well as the forward-backward correlation coefficient, 
will allow for a verification of production models of the early phase.
The awaited F-B correlation analysis with the LHC data will shed further light on the early production mechanism. The statistical 
method presented in this paper is directly applicable to that case and we have made one-parameter predictions 
for the dependence of the correlation coefficient on centrality at the LHC energies, applying the Glauber model.

\bigskip 

We thank Adam Bzdak for very useful discussions and pointing out the proper interpretation of the STAR data.
This research was supported by the Polish National Science Centre, grant DEC-2011/01/D/ST2/00772.

\appendix

\section{Superposition model \label{sec:super}}

In this Appendix we derive the relevant statistical formulas for the superposition model. 
Let the number of produced particles $n$ be composed of
independent emissions from $s$ sources, 
\begin{equation}
n=\sum_{i=1}^{s}m_i. \label{super}
\end{equation}
Here $m_i$ is the number of particles produced by the $i$th source from some universal distribution. 
Then the well-known relations follow: 
\begin{eqnarray}
\langle n \rangle &=& \langle s \rangle \langle m \rangle, \label{mean} \\
{\rm var}(n) &=& \langle s \rangle {\rm var}(m) + \langle m \rangle^2 {\rm var}(s) . \label{var} 
\end{eqnarray}
We give for completeness the derivation. Introduce
\begin{eqnarray}
\delta m_i = m_i - \langle m \rangle, \;\;\;{\rm with} \; \langle \delta m_i  \rangle = 0.
\label{shift}
\end{eqnarray}
Then 
\begin{eqnarray}
{\rm var}(n)
&=& \langle \sum_{i=1}^{s}(\delta m_i+\langle m \rangle)  
\sum_{j=1}^{s} (\delta m_j +  \langle m \rangle)\rangle - 
(\langle s \rangle \langle m \rangle)^2 \nonumber \\
&=& \langle \sum_{i=1}^{s} \delta m_i^2 \rangle + \langle \sum_{i,j=1,i \neq j}^{s} 
\delta m_i \delta m_j \rangle + 2 \langle m \rangle \langle \sum_{i=1}^{s}\delta m_i \rangle 
\nonumber \\
&+& \langle m \rangle^2 \langle \sum_{i=1}^{s} \sum_{j=1}^{s} \rangle -
\langle s \rangle^2 \langle m \rangle^2.
\end{eqnarray}
The second and third term in the last equality vanish due to Eq.~(\ref{shift}). Also,
from the independence of the production from different sources, the first term is
equal to $\langle s \rangle {\rm var}(m)$. Finally, using the obvious fact that
$\sum_{i=1}^{s} \sum_{j=1}^{s} =s^2$ we obtain 
Eq.~(\ref{var}). 

Next, we look at the covariance between two well-separated bins, which means $\langle m_i m_j \rangle = 
\langle m \rangle^2$, with $i$ and $j$ belonging to two different bins. 
We have 
\begin{eqnarray}
\langle n_1 n_2 \rangle = \langle \sum_{i=1}^{s_1} m_i  \sum_{j=1}^{s_2} m_j \rangle = 
\langle m \rangle^2 \langle s_1 s_2 \rangle, 
\end{eqnarray}
and 
\begin{eqnarray}
{\rm cov}(n_1,n_2) = \langle m \rangle^2 {\rm cov}(s_1,s_2).  \label{eq:cors}
\end{eqnarray}
For the correlation coefficient it follows that
\begin{eqnarray}
\rho(n_1,n_2)=\frac{\rho(s_1,s_2)}{\sqrt{1+ \frac{\omega(m)}{ \langle m \rangle\omega(s_1)}} \sqrt{1+ \frac{\omega(m)}{ \langle m \rangle\omega(s_2)}}}. \label{eq:fin}
\end{eqnarray}

\bibliographystyle{apsrev4-1-nohep}
\bibliography{hydro,hydr}

\begin{thebibliography}{10}%
\makeatletter
\providecommand \@ifxundefined [1]{%
 \ifx #1\undefined \expandafter \@firstoftwo
 \else \expandafter \@secondoftwo
\fi
}%
\providecommand \@ifnum [1]{%
 \ifnum #1\expandafter \@firstoftwo
 \else \expandafter \@secondoftwo
\fi
}%
\providecommand \enquote [1]{``#1''}%
\providecommand \bibnamefont  [1]{#1}%
\providecommand \bibfnamefont [1]{#1}%
\providecommand \citenamefont [1]{#1}%
\providecommand\href[0]{\@sanitize\@href}%
\providecommand\@href[1]{\endgroup\@@startlink{#1}\endgroup\@@href}%
\providecommand\@@href[1]{#1\@@endlink}%
\providecommand \@sanitize [0]{\begingroup\catcode`\&12\catcode`\#12\relax}%
\@ifxundefined \pdfoutput {\@firstoftwo}{%
 \@ifnum{\z@=\pdfoutput}{\@firstoftwo}{\@secondoftwo}%
}{%
 \providecommand\@@startlink[1]{\leavevmode\special{html:<a href="#1">}}%
 \providecommand\@@endlink[0]{\special{html:</a>}}%
}{%
 \providecommand\@@startlink[1]{%
  \leavevmode
  \pdfstartlink
   attr{/Border[0 0 1 ]/H/I/C[0 1 1]}%
   user{/Subtype/Link/A<</Type/Action/S/URI/URI(#1)>>}%
  \relax
 }%
 \providecommand\@@endlink[0]{\pdfendlink}%
}%
\providecommand \url  [0]{\begingroup\@sanitize \@url }%
\providecommand \@url [1]{\endgroup\@href {#1}{\urlprefix}}%
\providecommand \urlprefix [0]{URL }%
\providecommand \Eprint[0]{\href }%
\@ifxundefined \urlstyle {%
  \providecommand \doi [1]{doi:\discretionary{}{}{}#1}%
}{%
  \providecommand \doi [0]{doi:\discretionary{}{}{}\begingroup
  \urlstyle{rm}\Url }%
}%
\providecommand \doibase [0]{http://dx.doi.org/}%
\providecommand \Doi[1]{\href{\doibase#1}}%
\providecommand \bibAnnote [3]{%
  \BibitemShut{#1}%
  \begin{quotation}\noindent
    \textsc{Key:}\ #2\\\textsc{Annotation:}\ #3%
  \end{quotation}%
}%
\providecommand \bibAnnoteFile [2]{%
  \IfFileExists{#2}{\bibAnnote {#1} {#2} {\input{#2}}}{}%
}%
\providecommand \typeout [0]{\immediate \write \m@ne }%
\providecommand \selectlanguage [0]{\@gobble}%
\providecommand \bibinfo [0]{\@secondoftwo}%
\providecommand \bibfield [0]{\@secondoftwo}%
\providecommand \translation [1]{[#1]}%
\providecommand \BibitemOpen[0]{}%
\providecommand \bibitemStop [0]{}%
\providecommand \bibitemNoStop [0]{.\EOS\space}%
\providecommand \EOS [0]{\spacefactor3000\relax}%
\providecommand \BibitemShut [1]{\csname bibitem#1\endcsname}%
\bibitem{Back:2006id}%
  \BibitemOpen
  \bibfield{author}{%
  \bibinfo {author} {\bibfnamefont{B.}~\bibnamefont{Back}} \emph{et~al.}
  (\bibinfo {collaboration} {PHOBOS Collaboration}),\ }%
  \bibfield{journal}{%
  \Doi{10.1103/PhysRevC.74.011901}{\bibinfo {journal} {Phys.Rev.}}\ }%
  \textbf{\bibinfo {volume} {C74}},\ \bibinfo {pages} {011901} (\bibinfo {year}
  {2006})%
  \bibAnnoteFile{NoStop}{Back:2006id}%
\bibitem{Tarnowsky:2006nh}%
  \BibitemOpen
  \bibfield{author}{%
  \bibinfo {author} {\bibfnamefont{T.~J.}\ \bibnamefont{Tarnowsky}} (\bibinfo
  {collaboration} {STAR Collaboration})}%
   (\bibinfo {year} {2006}),\
  \Eprint{http://arxiv.org/abs/nucl-ex/0606018}{arXiv:nucl-ex/0606018
  [nucl-ex]}%
  \bibAnnoteFile{NoStop}{Tarnowsky:2006nh}%
\bibitem{Tarnowsky:2008am}%
  \BibitemOpen
  \bibfield{author}{%
  \bibinfo {author} {\bibfnamefont{T.~J.}\ \bibnamefont{Tarnowsky}}}%
   (\bibinfo {year} {2008}),\
  \Eprint{http://arxiv.org/abs/0807.1941}{arXiv:0807.1941 [nucl-ex]}%
  \bibAnnoteFile{NoStop}{Tarnowsky:2008am}%
\bibitem{Abelev:2009ag}%
  \BibitemOpen
  \bibfield{author}{%
  \bibinfo {author} {\bibfnamefont{B.}~\bibnamefont{Abelev}} \emph{et~al.}
  (\bibinfo {collaboration} {STAR Collaboration}),\ }%
  \bibfield{journal}{%
  \Doi{10.1103/PhysRevLett.103.172301}{\bibinfo {journal} {Phys.Rev.Lett.}}\ }%
  \textbf{\bibinfo {volume} {103}},\ \bibinfo {pages} {172301} (\bibinfo {year}
  {2009})%
  \bibAnnoteFile{NoStop}{Abelev:2009ag}%
\bibitem{Brogueira:2006yk}%
  \BibitemOpen
  \bibfield{author}{%
  \bibinfo {author} {\bibfnamefont{P.}~\bibnamefont{Brogueira}}\ and\ \bibinfo
  {author} {\bibfnamefont{J.}~\bibnamefont{Dias~de Deus}},\ }%
  \bibfield{journal}{%
  \Doi{10.1016/j.physletb.2007.01.076}{\bibinfo {journal} {Phys.Lett.}}\ }%
  \textbf{\bibinfo {volume} {B653}},\ \bibinfo {pages} {202} (\bibinfo {year}
  {2007})%
  \bibAnnoteFile{NoStop}{Brogueira:2006yk}%
\bibitem{Brogueira:2007ub}%
  \BibitemOpen
  \bibfield{author}{%
  \bibinfo {author} {\bibfnamefont{P.}~\bibnamefont{Brogueira}}, \bibinfo
  {author} {\bibfnamefont{J.}~\bibnamefont{Dias~de Deus}},\ and\ \bibinfo
  {author} {\bibfnamefont{J.~G.}\ \bibnamefont{Milhano}},\ }%
  \bibfield{journal}{%
  \Doi{10.1103/PhysRevC.76.064901}{\bibinfo {journal} {Phys.Rev.}}\ }%
  \textbf{\bibinfo {volume} {C76}},\ \bibinfo {pages} {064901} (\bibinfo {year}
  {2007})%
  \bibAnnoteFile{NoStop}{Brogueira:2007ub}%
\bibitem{Haussler:2006rg}%
  \BibitemOpen
  \bibfield{author}{%
  \bibinfo {author} {\bibfnamefont{S.}~\bibnamefont{Haussler}}, \bibinfo
  {author} {\bibfnamefont{M.}~\bibnamefont{Abdel-Aziz}},\ and\ \bibinfo
  {author} {\bibfnamefont{M.}~\bibnamefont{Bleicher}},\ }%
  \bibfield{journal}{%
  \Doi{10.1016/j.nuclphysa.2006.11.146}{\bibinfo {journal} {Nucl.Phys.}}\ }%
  \textbf{\bibinfo {volume} {A785}},\ \bibinfo {pages} {253} (\bibinfo {year}
  {2007})%
  \bibAnnoteFile{NoStop}{Haussler:2006rg}%
\bibitem{Bzdak:2009dr}%
  \BibitemOpen
  \bibfield{author}{%
  \bibinfo {author} {\bibfnamefont{A.}~\bibnamefont{Bzdak}}\ and\ \bibinfo
  {author} {\bibfnamefont{K.}~\bibnamefont{Wozniak}},\ }%
  \bibfield{journal}{%
  \Doi{10.1103/PhysRevC.81.034908}{\bibinfo {journal} {Phys.Rev.}}\ }%
  \textbf{\bibinfo {volume} {C81}},\ \bibinfo {pages} {034908} (\bibinfo {year}
  {2010})%
  \bibAnnoteFile{NoStop}{Bzdak:2009dr}%
\bibitem{Bzdak:2009xq}%
  \BibitemOpen
  \bibfield{author}{%
  \bibinfo {author} {\bibfnamefont{A.}~\bibnamefont{Bzdak}},\ }%
  \bibfield{journal}{%
  \Doi{10.1103/PhysRevC.80.024906}{\bibinfo {journal} {Phys.Rev.}}\ }%
  \textbf{\bibinfo {volume} {C80}},\ \bibinfo {pages} {024906} (\bibinfo {year}
  {2009})%
  \bibAnnoteFile{NoStop}{Bzdak:2009xq}%
\bibitem{Yan:2010et}%
  \BibitemOpen
  \bibfield{author}{%
  \bibinfo {author} {\bibfnamefont{Y.-L.}\ \bibnamefont{Yan}}, \bibinfo
  {author} {\bibfnamefont{D.-M.}\ \bibnamefont{Zhou}}, \bibinfo {author}
  {\bibfnamefont{B.-G.}\ \bibnamefont{Dong}}, \bibinfo {author}
  {\bibfnamefont{X.-M.}\ \bibnamefont{Li}}, \bibinfo {author}
  {\bibfnamefont{H.-L.}\ \bibnamefont{Ma}}, \emph{et~al.},\ }%
  \bibfield{journal}{%
  \Doi{10.1103/PhysRevC.81.044914}{\bibinfo {journal} {Phys.Rev.}}\ }%
  \textbf{\bibinfo {volume} {C81}},\ \bibinfo {pages} {044914} (\bibinfo {year}
  {2010})%
  \bibAnnoteFile{NoStop}{Yan:2010et}%
\bibitem{Lappi:2009vb}%
  \BibitemOpen
  \bibfield{author}{%
  \bibinfo {author} {\bibfnamefont{T.}~\bibnamefont{Lappi}}\ and\ \bibinfo
  {author} {\bibfnamefont{L.}~\bibnamefont{McLerran}},\ }%
  \bibfield{journal}{%
  \Doi{10.1016/j.nuclphysa.2009.11.003}{\bibinfo {journal} {Nucl.Phys.}}\ }%
  \textbf{\bibinfo {volume} {A832}},\ \bibinfo {pages} {330} (\bibinfo {year}
  {2010})%
  \bibAnnoteFile{NoStop}{Lappi:2009vb}%
\bibitem{Lappi:2010ek}%
  \BibitemOpen
  \bibfield{author}{%
  \bibinfo {author} {\bibfnamefont{T.}~\bibnamefont{Lappi}},\ }%
  \bibfield{journal}{%
  \Doi{10.1142/S0218301311017302}{\bibinfo {journal} {Int.J.Mod.Phys.}}\ }%
  \textbf{\bibinfo {volume} {E20}},\ \bibinfo {pages} {1} (\bibinfo {year}
  {2011})%
  \bibAnnoteFile{NoStop}{Lappi:2010ek}%
\bibitem{Bialas:2010zb}%
  \BibitemOpen
  \bibfield{author}{%
  \bibinfo {author} {\bibfnamefont{A.}~\bibnamefont{Bialas}}\ and\ \bibinfo
  {author} {\bibfnamefont{K.}~\bibnamefont{Zalewski}},\ }%
  \bibfield{journal}{%
  \Doi{10.1103/PhysRevC.85.029903, 10.1103/PhysRevC.82.034911}{\bibinfo
  {journal} {Phys.Rev.}}\ }%
  \textbf{\bibinfo {volume} {C82}},\ \bibinfo {pages} {034911} (\bibinfo {year}
  {2010})%
  \bibAnnoteFile{NoStop}{Bialas:2010zb}%
\bibitem{Bialas:2011bz}%
  \BibitemOpen
  \bibfield{author}{%
  \bibinfo {author} {\bibfnamefont{A.}~\bibnamefont{Bialas}}, \bibinfo {author}
  {\bibfnamefont{A.}~\bibnamefont{Bzdak}},\ and\ \bibinfo {author}
  {\bibfnamefont{K.}~\bibnamefont{Zalewski}},\ }%
  \bibfield{journal}{%
  \Doi{10.1016/j.physletb.2012.03.008}{\bibinfo {journal} {Phys.Lett.}}\ }%
  \textbf{\bibinfo {volume} {B710}},\ \bibinfo {pages} {332} (\bibinfo {year}
  {2012})%
  \bibAnnoteFile{NoStop}{Bialas:2011bz}%
\bibitem{Bialas:2011vj}%
  \BibitemOpen
  \bibfield{author}{%
  \bibinfo {author} {\bibfnamefont{A.}~\bibnamefont{Bialas}}\ and\ \bibinfo
  {author} {\bibfnamefont{K.}~\bibnamefont{Zalewski}},\ }%
  \bibfield{journal}{%
  \Doi{10.1016/j.physletb.2011.03.036}{\bibinfo {journal} {Phys.Lett.}}\ }%
  \textbf{\bibinfo {volume} {B698}},\ \bibinfo {pages} {416} (\bibinfo {year}
  {2011})%
  \bibAnnoteFile{NoStop}{Bialas:2011vj}%
\bibitem{Bialas:2011xk}%
  \BibitemOpen
  \bibfield{author}{%
  \bibinfo {author} {\bibfnamefont{A.}~\bibnamefont{Bialas}}\ and\ \bibinfo
  {author} {\bibfnamefont{K.}~\bibnamefont{Zalewski}},\ }%
  \bibfield{journal}{%
  \Doi{10.1016/j.nuclphysa.2011.05.006}{\bibinfo {journal} {Nucl.Phys.}}\ }%
  \textbf{\bibinfo {volume} {A860}},\ \bibinfo {pages} {56} (\bibinfo {year}
  {2011})%
  \bibAnnoteFile{NoStop}{Bialas:2011xk}%
\bibitem{Bzdak:2011nb}%
  \BibitemOpen
  \bibfield{author}{%
  \bibinfo {author} {\bibfnamefont{A.}~\bibnamefont{Bzdak}},\ }%
  \bibfield{journal}{%
  \Doi{10.1103/PhysRevC.85.051901}{\bibinfo {journal} {Phys.Rev.}}\ }%
  \textbf{\bibinfo {volume} {C85}},\ \bibinfo {pages} {051901} (\bibinfo {year}
  {2012})%
  \bibAnnoteFile{NoStop}{Bzdak:2011nb}%
\bibitem{Bzdak:2012tp}%
  \BibitemOpen
  \bibfield{author}{%
  \bibinfo {author} {\bibfnamefont{A.}~\bibnamefont{Bzdak}}\ and\ \bibinfo
  {author} {\bibfnamefont{D.}~\bibnamefont{Teaney}}}%
   (\bibinfo {year} {2012}),\
  \Eprint{http://arxiv.org/abs/1210.1965}{arXiv:1210.1965 [nucl-th]}%
  \bibAnnoteFile{NoStop}{Bzdak:2012tp}%
\bibitem{Fialkowski:2012ma}%
  \BibitemOpen
  \bibfield{author}{%
  \bibinfo {author} {\bibfnamefont{K.}~\bibnamefont{Fialkowski}}\ and\ \bibinfo
  {author} {\bibfnamefont{R.}~\bibnamefont{Wit}},\ }%
  \bibfield{journal}{%
  \Doi{10.2478/s11534-012-0109-9}{\bibinfo {journal} {Central Eur.J.Phys.}}\ }%
  \textbf{\bibinfo {volume} {10}},\ \bibinfo {pages} {1125} (\bibinfo {year}
  {2012})%
  \bibAnnoteFile{NoStop}{Fialkowski:2012ma}%
\bibitem{Bialas:1976ed}%
  \BibitemOpen
  \bibfield{author}{%
  \bibinfo {author} {\bibfnamefont{A.}~\bibnamefont{Bialas}}, \bibinfo {author}
  {\bibfnamefont{M.}~\bibnamefont{Bleszynski}},\ and\ \bibinfo {author}
  {\bibfnamefont{W.}~\bibnamefont{Czyz}},\ }%
  \bibfield{journal}{%
  \Doi{10.1016/0550-3213(76)90329-1}{\bibinfo {journal} {Nucl.Phys.}}\ }%
  \textbf{\bibinfo {volume} {B111}},\ \bibinfo {pages} {461} (\bibinfo {year}
  {1976})%
  \bibAnnoteFile{NoStop}{Bialas:1976ed}%
\bibitem{Bialas:2008zza}%
  \BibitemOpen
  \bibfield{author}{%
  \bibinfo {author} {\bibfnamefont{A.}~\bibnamefont{Bialas}},\ }%
  \bibfield{journal}{%
  \Doi{10.1088/0954-3899/35/4/044053}{\bibinfo {journal} {J.Phys.}}\ }%
  \textbf{\bibinfo {volume} {G35}},\ \bibinfo {pages} {044053} (\bibinfo {year}
  {2008})%
  \bibAnnoteFile{NoStop}{Bialas:2008zza}%
\bibitem{Broniowski:2007ft}%
  \BibitemOpen
  \bibfield{author}{%
  \bibinfo {author} {\bibfnamefont{W.}~\bibnamefont{Broniowski}}, \bibinfo
  {author} {\bibfnamefont{P.}~\bibnamefont{Bozek}},\ and\ \bibinfo {author}
  {\bibfnamefont{M.}~\bibnamefont{Rybczynski}},\ }%
  \bibfield{journal}{%
  \Doi{10.1103/PhysRevC.76.054905}{\bibinfo {journal} {Phys.Rev.}}\ }%
  \textbf{\bibinfo {volume} {C76}},\ \bibinfo {pages} {054905} (\bibinfo {year}
  {2007})%
  \bibAnnoteFile{NoStop}{Broniowski:2007ft}%
\bibitem{Kharzeev:2000ph}%
  \BibitemOpen
  \bibfield{author}{%
  \bibinfo {author} {\bibfnamefont{D.}~\bibnamefont{Kharzeev}}\ and\ \bibinfo
  {author} {\bibfnamefont{M.}~\bibnamefont{Nardi}},\ }%
  \bibfield{journal}{%
  \Doi{10.1016/S0370-2693(01)00457-9}{\bibinfo {journal} {Phys.Lett.}}\ }%
  \textbf{\bibinfo {volume} {B507}},\ \bibinfo {pages} {121} (\bibinfo {year}
  {2001})%
  \bibAnnoteFile{NoStop}{Kharzeev:2000ph}%
\bibitem{Back:2001xy}%
  \BibitemOpen
  \bibfield{author}{%
  \bibinfo {author} {\bibfnamefont{B.}~\bibnamefont{Back}} \emph{et~al.}
  (\bibinfo {collaboration} {PHOBOS Collaboration}),\ }%
  \bibfield{journal}{%
  \Doi{10.1103/PhysRevC.65.031901}{\bibinfo {journal} {Phys.Rev.}}\ }%
  \textbf{\bibinfo {volume} {C65}},\ \bibinfo {pages} {031901} (\bibinfo {year}
  {2002})%
  \bibAnnoteFile{NoStop}{Back:2001xy}%
\bibitem{Kovner:1995ja}%
  \BibitemOpen
  \bibfield{author}{%
  \bibinfo {author} {\bibfnamefont{A.}~\bibnamefont{Kovner}}, \bibinfo {author}
  {\bibfnamefont{L.~D.}\ \bibnamefont{McLerran}},\ and\ \bibinfo {author}
  {\bibfnamefont{H.}~\bibnamefont{Weigert}},\ }%
  \bibfield{journal}{%
  \Doi{10.1103/PhysRevD.52.6231}{\bibinfo {journal} {Phys.Rev.}}\ }%
  \textbf{\bibinfo {volume} {D52}},\ \bibinfo {pages} {6231} (\bibinfo {year}
  {1995})%
  \bibAnnoteFile{NoStop}{Kovner:1995ja}%
\bibitem{Iancu:2000hn}%
  \BibitemOpen
  \bibfield{author}{%
  \bibinfo {author} {\bibfnamefont{E.}~\bibnamefont{Iancu}}, \bibinfo {author}
  {\bibfnamefont{A.}~\bibnamefont{Leonidov}},\ and\ \bibinfo {author}
  {\bibfnamefont{L.~D.}\ \bibnamefont{McLerran}},\ }%
  \bibfield{journal}{%
  \Doi{10.1016/S0375-9474(01)00642-X}{\bibinfo {journal} {Nucl.Phys.}}\ }%
  \textbf{\bibinfo {volume} {A692}},\ \bibinfo {pages} {583} (\bibinfo {year}
  {2001})%
  \bibAnnoteFile{NoStop}{Iancu:2000hn}%
\bibitem{Lappi:2010cp}%
  \BibitemOpen
  \bibfield{author}{%
  \bibinfo {author} {\bibfnamefont{T.}~\bibnamefont{Lappi}},\ }%
  \bibfield{journal}{%
  \Doi{10.1143/PTPS.187.134}{\bibinfo {journal} {Prog.Theor.Phys.Suppl.}}\ }%
  \textbf{\bibinfo {volume} {187}},\ \bibinfo {pages} {134} (\bibinfo {year}
  {2011})%
  \bibAnnoteFile{NoStop}{Lappi:2010cp}%
\bibitem{Lappi:2011zz}%
  \BibitemOpen
  \bibfield{author}{%
  \bibinfo {author} {\bibfnamefont{T.}~\bibnamefont{Lappi}},\ }%
  \bibfield{journal}{%
  \Doi{10.1088/1742-6596/270/1/012055}{\bibinfo {journal} {J.Phys.Conf.Ser.}}\
  }%
  \textbf{\bibinfo {volume} {270}},\ \bibinfo {pages} {012055} (\bibinfo {year}
  {2011})%
  \bibAnnoteFile{NoStop}{Lappi:2011zz}%
\bibitem{Amelin:1994mf}%
  \BibitemOpen
  \bibfield{author}{%
  \bibinfo {author} {\bibfnamefont{N.}~\bibnamefont{Amelin}}, \bibinfo {author}
  {\bibfnamefont{N.}~\bibnamefont{Armesto}}, \bibinfo {author}
  {\bibfnamefont{M.}~\bibnamefont{Braun}}, \bibinfo {author}
  {\bibfnamefont{E.}~\bibnamefont{Ferreiro}},\ and\ \bibinfo {author}
  {\bibfnamefont{C.}~\bibnamefont{Pajares}},\ }%
  \bibfield{journal}{%
  \Doi{10.1103/PhysRevLett.73.2813}{\bibinfo {journal} {Phys.Rev.Lett.}}\ }%
  \textbf{\bibinfo {volume} {73}},\ \bibinfo {pages} {2813} (\bibinfo {year}
  {1994})%
  \bibAnnoteFile{NoStop}{Amelin:1994mf}%
\bibitem{Florkowski:2010zz}%
  \BibitemOpen
  \bibfield{author}{%
  \bibinfo {author} {\bibfnamefont{W.}~\bibnamefont{Florkowski}},\ }%
  \emph{\bibinfo {title} {{Phenomenology of Ultra-Relativistic Heavy-Ion
  Collisions}}}\ (\bibinfo {publisher} {World Scientific Publishing Company,
  Singapore},\ \bibinfo {year} {2010})%
  \bibAnnoteFile{NoStop}{Florkowski:2010zz}%
\bibitem{Broniowski:2007nz}%
  \BibitemOpen
  \bibfield{author}{%
  \bibinfo {author} {\bibfnamefont{W.}~\bibnamefont{Broniowski}}, \bibinfo
  {author} {\bibfnamefont{M.}~\bibnamefont{Rybczynski}},\ and\ \bibinfo
  {author} {\bibfnamefont{P.}~\bibnamefont{Bozek}},\ }%
  \bibfield{journal}{%
  \Doi{10.1016/j.cpc.2008.07.016}{\bibinfo {journal} {Comput.Phys.Commun.}}\ }%
  \textbf{\bibinfo {volume} {180}},\ \bibinfo {pages} {69} (\bibinfo {year}
  {2009})%
  \bibAnnoteFile{NoStop}{Broniowski:2007nz}%
\bibitem{Czyz:1969jg}%
  \BibitemOpen
  \bibfield{author}{%
  \bibinfo {author} {\bibfnamefont{W.}~\bibnamefont{Czyz}}\ and\ \bibinfo
  {author} {\bibfnamefont{L.}~\bibnamefont{Maximon}},\ }%
  \bibfield{journal}{%
  \Doi{10.1016/0003-4916(69)90321-2}{\bibinfo {journal} {Annals Phys.}}\ }%
  \textbf{\bibinfo {volume} {52}},\ \bibinfo {pages} {59} (\bibinfo {year}
  {1969})%
  \bibAnnoteFile{NoStop}{Czyz:1969jg}%
\bibitem{Kolb:2003dz}%
  \BibitemOpen
  \bibfield{author}{%
  \bibinfo {author} {\bibfnamefont{P.~F.}\ \bibnamefont{Kolb}}\ and\ \bibinfo
  {author} {\bibfnamefont{U.~W.}\ \bibnamefont{Heinz}},\ }%
  in\ \emph{\bibinfo {booktitle} {Quark Gluon Plasma 3}},\ \bibinfo {editor}
  {edited by\ \bibinfo {editor} {\bibfnamefont{R.}~\bibnamefont{Hwa}}\ and\
  \bibinfo {editor} {\bibfnamefont{X.~N.}\ \bibnamefont{Wang}}}\ (\bibinfo
  {publisher} {World Scientific, Singapore},\ \bibinfo {year} {2004})\
  \Eprint{http://arxiv.org/abs/nucl-th/0305084}{arXiv:nucl-th/0305084}%
  \bibAnnoteFile{NoStop}{Kolb:2003dz}%
\bibitem{Huovinen:2006jp}%
  \BibitemOpen
  \bibfield{author}{%
  \bibinfo {author} {\bibfnamefont{P.}~\bibnamefont{Huovinen}}\ and\ \bibinfo
  {author} {\bibfnamefont{P.~V.}\ \bibnamefont{Ruuskanen}},\ }%
  \bibfield{journal}{%
  \Doi{10.1146/annurev.nucl.54.070103.181236}{\bibinfo {journal} {Ann. Rev.
  Nucl. Part. Sci.}}\ }%
  \textbf{\bibinfo {volume} {56}},\ \bibinfo {pages} {163} (\bibinfo {year}
  {2006})%
  \bibAnnoteFile{NoStop}{Huovinen:2006jp}%
\bibitem{Andrade:2006yh}%
  \BibitemOpen
  \bibfield{author}{%
  \bibinfo {author} {\bibfnamefont{R.}~\bibnamefont{Andrade}}, \bibinfo
  {author} {\bibfnamefont{F.}~\bibnamefont{Grassi}}, \bibinfo {author}
  {\bibfnamefont{Y.}~\bibnamefont{Hama}}, \bibinfo {author}
  {\bibfnamefont{T.}~\bibnamefont{Kodama}},\ and\ \bibinfo {author}
  {\bibfnamefont{J.}~\bibnamefont{Socolowski}, \bibfnamefont{O.}},\ }%
  \bibfield{journal}{%
  \bibinfo {journal} {Phys. Rev. Lett.}\ }%
  \textbf{\bibinfo {volume} {97}},\ \bibinfo {pages} {202302} (\bibinfo {year}
  {2006})%
  \bibAnnoteFile{NoStop}{Andrade:2006yh}%
\bibitem{Werner:2009fa}%
  \BibitemOpen
  \bibfield{author}{%
  \bibinfo {author} {\bibfnamefont{K.}~\bibnamefont{Werner}} \emph{et~al.},\ }%
  \bibfield{journal}{%
  \bibinfo {journal} {J. Phys.}\ }%
  \textbf{\bibinfo {volume} {G36}},\ \bibinfo {pages} {064030} (\bibinfo {year}
  {2009})%
  \bibAnnoteFile{NoStop}{Werner:2009fa}%
\bibitem{Petersen:2010cw}%
  \BibitemOpen
  \bibfield{author}{%
  \bibinfo {author} {\bibfnamefont{H.}~\bibnamefont{Petersen}}, \bibinfo
  {author} {\bibfnamefont{G.-Y.}\ \bibnamefont{Qin}}, \bibinfo {author}
  {\bibfnamefont{S.~A.}\ \bibnamefont{Bass}},\ and\ \bibinfo {author}
  {\bibfnamefont{B.}~\bibnamefont{Muller}},\ }%
  \bibfield{journal}{%
  \Doi{10.1103/PhysRevC.82.041901}{\bibinfo {journal} {Phys.Rev.}}\ }%
  \textbf{\bibinfo {volume} {C82}},\ \bibinfo {pages} {041901} (\bibinfo {year}
  {2010})%
  \bibAnnoteFile{NoStop}{Petersen:2010cw}%
\bibitem{Holopainen:2010gz}%
  \BibitemOpen
  \bibfield{author}{%
  \bibinfo {author} {\bibfnamefont{H.}~\bibnamefont{Holopainen}}, \bibinfo
  {author} {\bibfnamefont{H.}~\bibnamefont{Niemi}},\ and\ \bibinfo {author}
  {\bibfnamefont{K.~J.}\ \bibnamefont{Eskola}},\ }%
  \bibfield{journal}{%
  \Doi{10.1103/PhysRevC.83.034901}{\bibinfo {journal} {Phys.Rev.}}\ }%
  \textbf{\bibinfo {volume} {C83}},\ \bibinfo {pages} {034901} (\bibinfo {year}
  {2011})%
  \bibAnnoteFile{NoStop}{Holopainen:2010gz}%
\bibitem{Gardim:2011xv}%
  \BibitemOpen
  \bibfield{author}{%
  \bibinfo {author} {\bibfnamefont{F.~G.}\ \bibnamefont{Gardim}}, \bibinfo
  {author} {\bibfnamefont{F.}~\bibnamefont{Grassi}}, \bibinfo {author}
  {\bibfnamefont{M.}~\bibnamefont{Luzum}},\ and\ \bibinfo {author}
  {\bibfnamefont{J.-Y.}\ \bibnamefont{Ollitrault}}}%
   (\bibinfo {year} {2011}),\
  \Eprint{http://arxiv.org/abs/1111.6538}{arXiv:1111.6538 [nucl-th]}%
  \bibAnnoteFile{NoStop}{Gardim:2011xv}%
\bibitem{Bozek:2011if}%
  \BibitemOpen
  \bibfield{author}{%
  \bibinfo {author} {\bibfnamefont{P.}~\bibnamefont{Bozek}},\ }%
  \bibfield{journal}{%
  \bibinfo {journal} {Phys.Rev.}\ }%
  \textbf{\bibinfo {volume} {C85}},\ \bibinfo {pages} {014911} (\bibinfo {year}
  {2012})%
  \bibAnnoteFile{NoStop}{Bozek:2011if}%
\bibitem{Schenke:2010rr}%
  \BibitemOpen
  \bibfield{author}{%
  \bibinfo {author} {\bibfnamefont{B.}~\bibnamefont{Schenke}}, \bibinfo
  {author} {\bibfnamefont{S.}~\bibnamefont{Jeon}},\ and\ \bibinfo {author}
  {\bibfnamefont{C.}~\bibnamefont{Gale}},\ }%
  \bibfield{journal}{%
  \Doi{10.1103/PhysRevLett.106.042301}{\bibinfo {journal} {Phys. Rev. Lett.}}\
  }%
  \textbf{\bibinfo {volume} {106}},\ \bibinfo {pages} {042301} (\bibinfo {year}
  {2011})%
  \bibAnnoteFile{NoStop}{Schenke:2010rr}%
\bibitem{Qiu:2011fi}%
  \BibitemOpen
  \bibfield{author}{%
  \bibinfo {author} {\bibfnamefont{Z.}~\bibnamefont{Qiu}}\ and\ \bibinfo
  {author} {\bibfnamefont{U.~W.}\ \bibnamefont{Heinz}}}%
   (\bibinfo {year} {2011}),\
  \Eprint{http://arxiv.org/abs/1108.1714}{arXiv:1108.1714 [nucl-th]}%
  \bibAnnoteFile{NoStop}{Qiu:2011fi}%
\bibitem{Chaudhuri:2011pa}%
  \BibitemOpen
  \bibfield{author}{%
  \bibinfo {author} {\bibfnamefont{A.}~\bibnamefont{Chaudhuri}}}%
   (\bibinfo {year} {2011}),\
  \Eprint{http://arxiv.org/abs/1112.1166}{arXiv:1112.1166 [nucl-th]}%
  \bibAnnoteFile{NoStop}{Chaudhuri:2011pa}%
\bibitem{Bozek:2012en}%
  \BibitemOpen
  \bibfield{author}{%
  \bibinfo {author} {\bibfnamefont{P.}~\bibnamefont{Bozek}}\ and\ \bibinfo
  {author} {\bibfnamefont{W.}~\bibnamefont{Broniowski}},\ }%
  \bibfield{journal}{%
  \Doi{10.1103/PhysRevLett.109.062301}{\bibinfo {journal} {Phys.Rev.Lett.}}\ }%
  \textbf{\bibinfo {volume} {109}},\ \bibinfo {pages} {062301} (\bibinfo {year}
  {2012})%
  \bibAnnoteFile{NoStop}{Bozek:2012en}%
\bibitem{Cooper:1974mv}%
  \BibitemOpen
  \bibfield{author}{%
  \bibinfo {author} {\bibfnamefont{F.}~\bibnamefont{Cooper}}\ and\ \bibinfo
  {author} {\bibfnamefont{G.}~\bibnamefont{Frye}},\ }%
  \bibfield{journal}{%
  \Doi{10.1103/PhysRevD.10.186}{\bibinfo {journal} {Phys. Rev.}}\ }%
  \textbf{\bibinfo {volume} {D10}},\ \bibinfo {pages} {186} (\bibinfo {year}
  {1974})%
  \bibAnnoteFile{NoStop}{Cooper:1974mv}%
\bibitem{BraunMunzinger:2001ip}%
  \BibitemOpen
  \bibfield{author}{%
  \bibinfo {author} {\bibfnamefont{P.}~\bibnamefont{Braun-Munzinger}}, \bibinfo
  {author} {\bibfnamefont{D.}~\bibnamefont{Magestro}}, \bibinfo {author}
  {\bibfnamefont{K.}~\bibnamefont{Redlich}},\ and\ \bibinfo {author}
  {\bibfnamefont{J.}~\bibnamefont{Stachel}},\ }%
  \bibfield{journal}{%
  \Doi{10.1016/S0370-2693(01)01069-3}{\bibinfo {journal} {Phys.Lett.}}\ }%
  \textbf{\bibinfo {volume} {B518}},\ \bibinfo {pages} {41} (\bibinfo {year}
  {2001})%
  \bibAnnoteFile{NoStop}{BraunMunzinger:2001ip}%
\bibitem{Broniowski:2001we}%
  \BibitemOpen
  \bibfield{author}{%
  \bibinfo {author} {\bibfnamefont{W.}~\bibnamefont{Broniowski}}\ and\ \bibinfo
  {author} {\bibfnamefont{W.}~\bibnamefont{Florkowski}},\ }%
  \bibfield{journal}{%
  \Doi{10.1103/PhysRevLett.87.272302}{\bibinfo {journal} {Phys.Rev.Lett.}}\ }%
  \textbf{\bibinfo {volume} {87}},\ \bibinfo {pages} {272302} (\bibinfo {year}
  {2001})%
  \bibAnnoteFile{NoStop}{Broniowski:2001we}%
\bibitem{Torrieri:2002jp}%
  \BibitemOpen
  \bibfield{author}{%
  \bibinfo {author} {\bibfnamefont{G.}~\bibnamefont{Torrieri}}\ and\ \bibinfo
  {author} {\bibfnamefont{J.}~\bibnamefont{Rafelski}},\ }%
  \bibfield{journal}{%
  \Doi{10.1103/PhysRevC.68.034912}{\bibinfo {journal} {Phys.Rev.}}\ }%
  \textbf{\bibinfo {volume} {C68}},\ \bibinfo {pages} {034912} (\bibinfo {year}
  {2003})%
  \bibAnnoteFile{NoStop}{Torrieri:2002jp}%
\bibitem{Rafelski:2003ju}%
  \BibitemOpen
  \bibfield{author}{%
  \bibinfo {author} {\bibfnamefont{J.}~\bibnamefont{Rafelski}}\ and\ \bibinfo
  {author} {\bibfnamefont{J.}~\bibnamefont{Letessier}},\ }%
  \bibfield{journal}{%
  \bibinfo {journal} {Acta Phys.Polon.}\ }%
  \textbf{\bibinfo {volume} {B34}},\ \bibinfo {pages} {5791} (\bibinfo {year}
  {2003})%
  \bibAnnoteFile{NoStop}{Rafelski:2003ju}%
\bibitem{Becattini:2003wp}%
  \BibitemOpen
  \bibfield{author}{%
  \bibinfo {author} {\bibfnamefont{F.}~\bibnamefont{Becattini}}, \bibinfo
  {author} {\bibfnamefont{M.}~\bibnamefont{Gazdzicki}}, \bibinfo {author}
  {\bibfnamefont{A.}~\bibnamefont{Keranen}}, \bibinfo {author}
  {\bibfnamefont{J.}~\bibnamefont{Manninen}},\ and\ \bibinfo {author}
  {\bibfnamefont{R.}~\bibnamefont{Stock}},\ }%
  \bibfield{journal}{%
  \Doi{10.1103/PhysRevC.69.024905}{\bibinfo {journal} {Phys.Rev.}}\ }%
  \textbf{\bibinfo {volume} {C69}},\ \bibinfo {pages} {024905} (\bibinfo {year}
  {2004})%
  \bibAnnoteFile{NoStop}{Becattini:2003wp}%
\bibitem{Torrieri:2004zz}%
  \BibitemOpen
  \bibfield{author}{%
  \bibinfo {author} {\bibfnamefont{G.}~\bibnamefont{Torrieri}}, \bibinfo
  {author} {\bibfnamefont{S.}~\bibnamefont{Steinke}}, \bibinfo {author}
  {\bibfnamefont{W.}~\bibnamefont{Broniowski}}, \bibinfo {author}
  {\bibfnamefont{W.}~\bibnamefont{Florkowski}}, \bibinfo {author}
  {\bibfnamefont{J.}~\bibnamefont{Letessier}}, \emph{et~al.},\ }%
  \bibfield{journal}{%
  \Doi{10.1016/j.cpc.2005.01.004}{\bibinfo {journal} {Comput.Phys.Commun.}}\ }%
  \textbf{\bibinfo {volume} {167}},\ \bibinfo {pages} {229} (\bibinfo {year}
  {2005})%
  \bibAnnoteFile{NoStop}{Torrieri:2004zz}%
\bibitem{Wheaton:2004qb}%
  \BibitemOpen
  \bibfield{author}{%
  \bibinfo {author} {\bibfnamefont{S.}~\bibnamefont{Wheaton}}\ and\ \bibinfo
  {author} {\bibfnamefont{J.}~\bibnamefont{Cleymans}},\ }%
  \bibfield{journal}{%
  \Doi{10.1016/j.cpc.2008.08.001}{\bibinfo {journal} {Comput.Phys.Commun.}}\ }%
  \textbf{\bibinfo {volume} {180}},\ \bibinfo {pages} {84} (\bibinfo {year}
  {2009})%
  \bibAnnoteFile{NoStop}{Wheaton:2004qb}%
\bibitem{Kisiel:2005hn}%
  \BibitemOpen
  \bibfield{author}{%
  \bibinfo {author} {\bibfnamefont{A.}~\bibnamefont{Kisiel}}, \bibinfo {author}
  {\bibfnamefont{T.}~\bibnamefont{{Ta\l{}u\'c}}}, \bibinfo {author}
  {\bibfnamefont{W.}~\bibnamefont{Broniowski}},\ and\ \bibinfo {author}
  {\bibfnamefont{W.}~\bibnamefont{Florkowski}},\ }%
  \bibfield{journal}{%
  \bibinfo {journal} {Comput. Phys. Commun.}\ }%
  \textbf{\bibinfo {volume} {174}},\ \bibinfo {pages} {669} (\bibinfo {year}
  {2006})%
  \bibAnnoteFile{NoStop}{Kisiel:2005hn}%
\bibitem{Amelin:2006qe}%
  \BibitemOpen
  \bibfield{author}{%
  \bibinfo {author} {\bibfnamefont{N.}~\bibnamefont{Amelin}}, \bibinfo {author}
  {\bibfnamefont{R.}~\bibnamefont{Lednicky}}, \bibinfo {author}
  {\bibfnamefont{T.}~\bibnamefont{Pocheptsov}}, \bibinfo {author}
  {\bibfnamefont{I.}~\bibnamefont{Lokhtin}}, \bibinfo {author}
  {\bibfnamefont{L.}~\bibnamefont{Malinina}}, \emph{et~al.},\ }%
  \bibfield{journal}{%
  \Doi{10.1103/PhysRevC.74.064901}{\bibinfo {journal} {Phys.Rev.}}\ }%
  \textbf{\bibinfo {volume} {C74}},\ \bibinfo {pages} {064901} (\bibinfo {year}
  {2006})%
  \bibAnnoteFile{NoStop}{Amelin:2006qe}%
\bibitem{Tomasik:2008fq}%
  \BibitemOpen
  \bibfield{author}{%
  \bibinfo {author} {\bibfnamefont{B.}~\bibnamefont{Tomasik}},\ }%
  \bibfield{journal}{%
  \Doi{10.1016/j.cpc.2009.02.019}{\bibinfo {journal} {Comput.Phys.Commun.}}\ }%
  \textbf{\bibinfo {volume} {180}},\ \bibinfo {pages} {1642} (\bibinfo {year}
  {2009})%
  \bibAnnoteFile{NoStop}{Tomasik:2008fq}%
\bibitem{Chojnacki:2011hb}%
  \BibitemOpen
  \bibfield{author}{%
  \bibinfo {author} {\bibfnamefont{M.}~\bibnamefont{Chojnacki}}, \bibinfo
  {author} {\bibfnamefont{A.}~\bibnamefont{Kisiel}}, \bibinfo {author}
  {\bibfnamefont{W.}~\bibnamefont{Florkowski}},\ and\ \bibinfo {author}
  {\bibfnamefont{W.}~\bibnamefont{Broniowski}},\ }%
  \bibfield{journal}{%
  \Doi{10.1016/j.cpc.2011.11.018}{\bibinfo {journal} {Comput. Phys. Commun.}}\
  }%
  \textbf{\bibinfo {volume} {183}},\ \bibinfo {pages} {746} (\bibinfo {year}
  {2012})%
  \bibAnnoteFile{NoStop}{Chojnacki:2011hb}%
\bibitem{Broniowski:2008vp}%
  \BibitemOpen
  \bibfield{author}{%
  \bibinfo {author} {\bibfnamefont{W.}~\bibnamefont{Broniowski}}, \bibinfo
  {author} {\bibfnamefont{M.}~\bibnamefont{Chojnacki}}, \bibinfo {author}
  {\bibfnamefont{W.}~\bibnamefont{Florkowski}},\ and\ \bibinfo {author}
  {\bibfnamefont{A.}~\bibnamefont{Kisiel}},\ }%
  \bibfield{journal}{%
  \Doi{10.1103/PhysRevLett.101.022301}{\bibinfo {journal} {Phys.Rev.Lett.}}\ }%
  \textbf{\bibinfo {volume} {101}},\ \bibinfo {pages} {022301} (\bibinfo {year}
  {2008})%
  \bibAnnoteFile{NoStop}{Broniowski:2008vp}%
\bibitem{Kapusta:2011gt}%
  \BibitemOpen
  \bibfield{author}{%
  \bibinfo {author} {\bibfnamefont{J.}~\bibnamefont{Kapusta}}, \bibinfo
  {author} {\bibfnamefont{B.}~\bibnamefont{Muller}},\ and\ \bibinfo {author}
  {\bibfnamefont{M.}~\bibnamefont{Stephanov}},\ }%
  \bibfield{journal}{%
  \Doi{10.1103/PhysRevC.85.054906}{\bibinfo {journal} {Phys.Rev.}}\ }%
  \textbf{\bibinfo {volume} {C85}},\ \bibinfo {pages} {054906} (\bibinfo {year}
  {2012})%
  \bibAnnoteFile{NoStop}{Kapusta:2011gt}%
\bibitem{Bozek:2011ua}%
  \BibitemOpen
  \bibfield{author}{%
  \bibinfo {author} {\bibfnamefont{P.}~\bibnamefont{Bozek}},\ }%
  \bibfield{journal}{%
  \Doi{10.1103/PhysRevC.85.034901}{\bibinfo {journal} {Phys.Rev.}}\ }%
  \textbf{\bibinfo {volume} {C85}},\ \bibinfo {pages} {034901} (\bibinfo {year}
  {2012})%
  \bibAnnoteFile{NoStop}{Bozek:2011ua}%
\bibitem{Abelev:2008ab}%
  \BibitemOpen
  \bibfield{author}{%
  \bibinfo {author} {\bibfnamefont{B.}~\bibnamefont{Abelev}} \emph{et~al.}
  (\bibinfo {collaboration} {STAR Collaboration}),\ }%
  \bibfield{journal}{%
  \Doi{10.1103/PhysRevC.79.034909}{\bibinfo {journal} {Phys.Rev.}}\ }%
  \textbf{\bibinfo {volume} {C79}},\ \bibinfo {pages} {034909} (\bibinfo {year}
  {2009})%
  \bibAnnoteFile{NoStop}{Abelev:2008ab}%
\end{thebibliography}%

\end{document}